\documentclass[a4paper,11pt]{article}
\pdfoutput=1 
\usepackage{jheppub} 
\usepackage{epsfig}
\usepackage{caption}
\usepackage{subcaption}
\usepackage{color,soul}
\usepackage{float}
\usepackage{amsfonts}
\usepackage{url}
\usepackage{slashed}
\usepackage{accents}
\usepackage{lipsum}
\usepackage{dirtytalk}
\usepackage{mathtools}
\usepackage{physics}
\usepackage[normalem]{ulem}

\title{Asymmetric Cannibal Dark Matter: Constraints from Neutron Star}
\author[]{Ujjal Kumar Dey,}
\author[]{Sourav Gope}
\affiliation[]{Department of Physical Sciences, Indian Institute of Science Education and Research Berhampur, \\
Ganjam, Odisha 760003, India}
\emailAdd{ujjal@iiserbpr.ac.in}
\emailAdd{souravg@iiserbpr.ac.in}

\abstract{Asymmetric dark matter can be efficiently captured by neutron stars  via elastic scattering with nucleons and dark matter self scattering. The accumulated dark matter thermalizes and concentrates in the stellar interior, forming a dark matter core. In this work, we propose a novel framework in which a $\mathbb{Z}_3$ symmetry allows for number-changing self-interactions of the form $3 \rightarrow 2$ within the dark sector. These cannibalistic reactions become increasingly efficient at high dark matter densities, leading to a significant depletion of the dark matter population in the stellar core. This number depletion heats up the neutron star above the standard cooling expectations, yielding observable thermal signatures in relatively old, isolated neutron stars, potentially detectable via future infrared telescopes. We show that even in the presence of other heating mechanisms, e.g. dark matter annihilation and kinetic heating, the cannibal heating dominates for certain parameter space. We demonstrate that the cannibal heating can predict detectable heating signatures in old neutron stars, thereby allowing a broader range of viable dark matter masses and couplings to the Standard Model.}

\allowdisplaybreaks

\begin{document}
\maketitle
\flushbottom

%%%%%%%%%%%%%%%%%%%%%%%%%%%%%%
\section{Introduction}
\label{sec:Introduction}
%%%%%%%%%%%%%%%%%%%%%%%%%%%%%%
Dark matter (DM) provides a compelling explanation for a range of astrophysical and cosmological observations, including galaxy rotation curves, large-scale structure formation, and the dynamics of galaxy clusters such as the Bullet Cluster. While its gravitational effects are well established, the particle nature of dark matter remains one of the key open questions in modern physics. In addition to direct and indirect searches conducted via terrestrial experiments, compact astrophysical objects such as neutron stars (NSs), white dwarfs offer promising avenues to probe DM interactions. Due to their extreme densities and strong gravitational fields, neutron stars can efficiently capture DM particles via elastic scattering with nucleons~\cite{Guver:2012ba}. As a result, NSs can serve as natural laboratories for testing DM-Standard Model (SM) interactions~\cite{Raj:2017wrv, Dasgupta:2020dik, Nguyen:2022zwb, Linden:2024uph, Maity:2021fxw}. Once captured, DM particles can deposit energy into the stellar interior through repeated scattering or, in some scenarios, via annihilation. In case where the annihilation to SM is kinematically forbidden, then also the DM may undergo annihilation aided by massive SM-like particles \cite{Dey:2016qgf}. These heating effects can raise the temperature of the neutron star above the standard cooling expectations and are sensitive to the DM mass and interaction strength~\cite{Ellis:2018bkr, Chen:2018ohx, Bramante:2013hn, Ema:2024wqr}. Observations of cold, old NSs can therefore provide complementary constraints on DM properties. 
In the case of asymmetric dark matter (ADM), where the relic abundance arises from a particle-antiparticle asymmetry, the DM is typically modeled as a complex scalar or fermion, and does not self-annihilate~\cite{Zurek:2013wia}. After capture, ADM can thermalize with the NS matter~\cite{Bertoni:2013bsa,Garani:2020wge} and accumulate in the stellar core. If the number of captured particles exceeds a critical threshold, the ADM core may undergo gravitational collapse to form a black hole (BH)~\cite{McDermott:2011jp}, which could eventually consume the entire star. The observed survival of neutron stars thus places stringent upper bounds on the DM mass and DM--SM interaction cross-section~\cite{Bell:2013xk,Garani:2018kkd, Bhattacharya:2023stq, Dutta:2024vzw}. 
Usually, ADM is motivated by the observed matter-antimatter asymmetry in the visible sector, suggesting the possibility of a shared origin for both baryonic and dark matter abundances. In scenarios where the dark and visible sectors are connected through a common mechanism in the early Universe, a CP-violating process can generate particle-antiparticle asymmetries in both sectors~\cite{Petraki:2013wwa}. In such frameworks, the dark sector may exhibit a gauge symmetric structure analogous to that of the SM, leading to the so-called mirror dark sector models~\cite{Foot:2013msa}. 
One class of models proposes that the asymmetry originates in the dark sector, possibly via a first-order phase transition accompanied by CP violation, and is subsequently transferred to the visible sector, thereby producing the baryon asymmetry. This mechanism is referred to as \textit{darkogenesis}~\cite{Shelton:2010ta}. Conversely, an initial baryon-lepton asymmetry generated in the visible sector can also be communicated to the dark sector, provided the dark matter is charged under \textit{$(B-L)$}, resulting in a mirrored asymmetry \cite{Zurek:2013wia}. More recent studies have shown that DM ($\chi$) semi-annihilation processes of the form $\chi\chi \to \chi^{\dagger}\phi$, where $\phi$ mixes with or decays into SM particles, can also generate an asymmetry within the dark sector itself \cite{Ghosh:2020lma}.
In this work, we consider a complex scalar ADM candidate stabilized by a $\mathbb{Z}_3$ symmetry, which naturally permits number-changing self-interactions of the form $3 \to 2$. These cannibalistic reactions become more efficient at high DM densities, leading to a depletion of the DM number density in the stellar core. Within the framework of a complex scalar DM, there can be an initial particle-antiparticle number asymmetry which can be present in the DM halo of a galaxy. Depending on the production mechanism, and interactions involving the DM sector, the evolution of this asymmetry can be different \cite{Cervantes:2024ipg,Hektor:2019ote}. In this work, we show that even for a completely asymmetric DM (i.e., with vanishing number density of antiparticles), the $3\rightarrow2$ reactions can generate a non-zero antiparticle content. With this phenomena, the DM annihilations come into picture and this makes the scenario different from the existing literature. 
In addition to depleting the DM number, the $3 \rightarrow 2$ processes inject kinetic energy into the DM particles, effectively heating the dark sector. This energy is subsequently transferred to the NS interior via DM-nucleon scatterings, establishing a thermal equilibrium temperature that depends on the DM properties. As a result, the NS temperature deviates from standard cooling mechanisms in the absence of DM.
Heating of neutron stars induced by DM annihilation into SM particles can dominate over the kinetic heating contribution~\cite{Keung:2020teb}. In our analysis, we demonstrate that even for large annihilation cross sections, this channel can be strongly suppressed if the initial abundance of DM antiparticles is constrained to be small. Consequently, in scenarios with a sizeable initial particle-antiparticle asymmetry, the annihilation probability is significantly reduced, leading to a correspondingly smaller heating effect. We identify the region of parameter space where cannibal heating dominates over kinetic heating caused by DM-nucleon elastic scatterings and heating by DM annihilation to SM particles during the capture process. While this heating mechanism is negligible for young neutron stars due to their high background temperatures, it becomes significant in older, isolated NSs. In particular, we find that for certain choices of DM mass and couplings, the resulting surface temperatures can reach $\mathcal{O}(1000)$~K which lie within the sensitivity range of near-infrared instruments aboard the James Webb Space Telescope (JWST), offering a promising observational window into the thermal signatures of cannibal dark matter.
%

%\vspace{0.5cm}
%
This article is organized as follows. In Sec.~\ref{sec:Evolution_DM_number}, we introduce the DM set-up and describe the evolution of the DM number inside neutron stars. The evolution of the neutron star temperature, incorporating cannibal, annihilation, and kinetic heating effects, is presented in Sec.~\ref{sec:Evolution_NS_temperature}. We discuss the observable surface temperatures of old neutron stars and the potential detectability of such thermal signals by the JWST and other upcoming telescopes in Sec.~\ref{sec:Observation_NS_temperature}. Finally, we conclude with a summary of our results and a discussion of future prospects.

%%%%%%%%%%%%%%%%%%%%%%%%%%%%%%
\section{Evolution of DM number inside NS}
\label{sec:Evolution_DM_number}
%%%%%%%%%%%%%%%%%%%%%%%%%%%%%%
We consider a scenario where an asymmetric scalar dark matter ($\chi$) with  strong self-interaction, is stabilized by a $\mathbb{Z}_3$ symmetry. The relevant Lagrangian can be given as,
\begin{align}
\mathcal{L}_{\text{int}}\supset m_\chi^2|\chi|^2 + \frac{\mu}{3!}\left(\chi^3+(\chi^\dagger)^3\right) + \frac{\lambda}{2!2!}\left(\chi^\dagger\chi\right)^2.
\label{eq:Lagrangian}
\end{align}
The $\mathbb{Z}_3$ symmetry enables a family of $3\rightarrow2$ as well as $2\rightarrow2$ DM self interactions. In addition to the self interactions, we also assume scattering and annihilation of DM with the SM particles, namely, $\chi n\rightarrow\chi n $, $\chi\chi^\dagger\rightarrow \text{SM}+\text{SM}$, where \textit{n} denotes the neutron. Here we remain agnostic about the underlying model or portal to the visible sector. We now discuss the possible Feynman diagrams, relevant cross sections for the $3\to 2$ processes, and show the contribution of each of the interaction in determining the evolution of DM within a NS.
\subsection{DM 3$\rightarrow$2 interactions}
The trilinear and quartic interaction enables $3\rightarrow 2$ type of cannibal processes which converts the mass-energy of the initial state to the kinetic energy of final state particles, if it occurs in the non-relativistic regime. The phenomenology of cannibal dark matter and cosmological bounds on the DM mass and self-couplings has been well studied in~\cite{Ghosh:2022asg}. 
\begin{figure}[hbtp]
\centering
\includegraphics[scale=0.45]{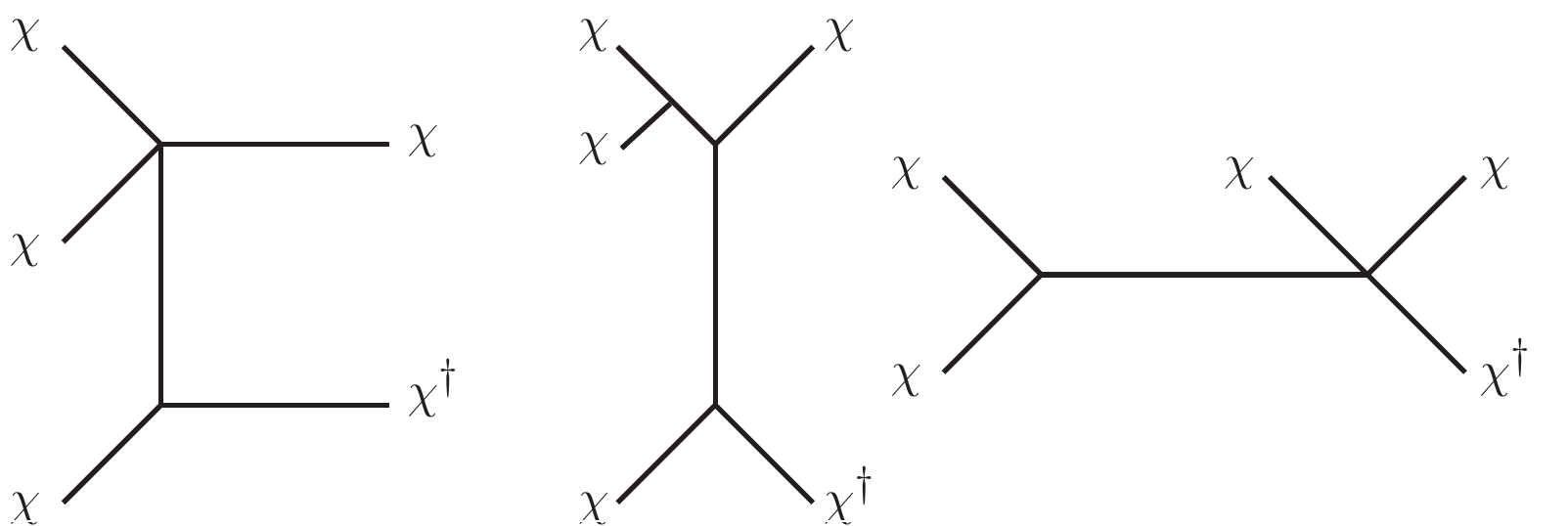}

\vspace{0.4cm}

\includegraphics[scale=0.55]{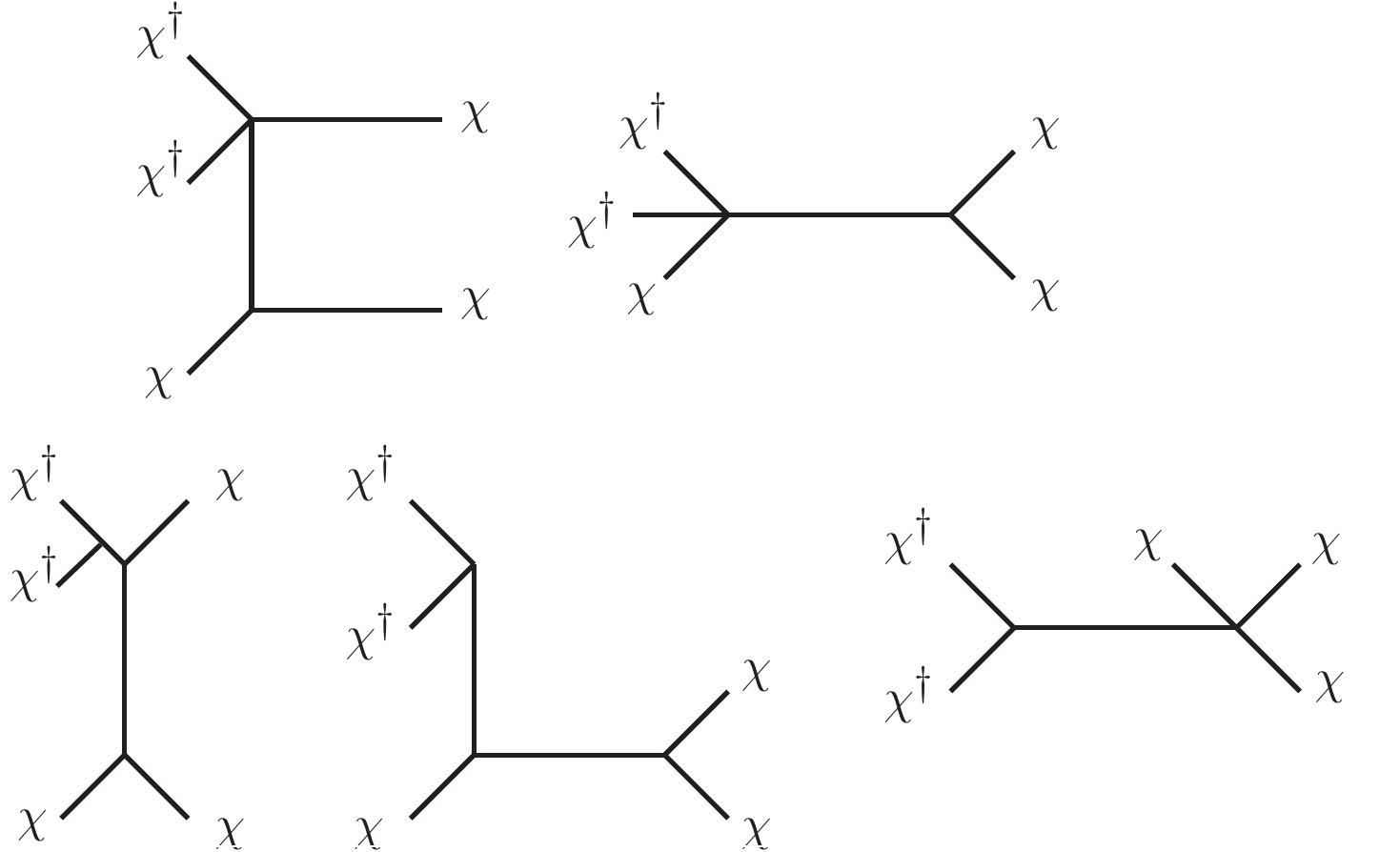}
\caption{\small{Possible diagrams for $3\rightarrow 2$ self-scatterings where DM is a complex scalar stabilized by a $\mathbb{Z}_3$ symmetry. The upper set of 3 diagrams indicate the process $\chi\chi\chi\rightarrow\chi\chi^{\dagger}$ while the rest of the others represent  $\chi\chi^{\dagger}\chi^{\dagger}\rightarrow\chi\chi$ interactions.}}
\label{fig:3to2_diagrams}
\end{figure}
The $\mathbb{Z}_3$ symmetry allows processes with both single and double propagator diagrams. The double propagator diagrams involve only the trilinear coupling $\mu$ whereas the diagram with the single propagator involves both trilinear ($\mu$) and quartic coupling ($\lambda$). The possible $3\rightarrow2$ processes respecting the $\mathbb{Z}_3$ symmetry are -- (i) $\chi\chi\chi\rightarrow\chi\chi^{\dagger}$, (ii) $\chi\chi^{\dagger}\chi^{\dagger}\rightarrow\chi\chi$ and their conjugate processes. The total matrix element combining all the diagrams for these two classes of processes in Fig.~\ref{fig:3to2_diagrams}, including all the permutations of momenta are,
\begin{align}
\mathcal{M}_{\chi\chi\chi\rightarrow\chi\chi^{\dagger}} &=\frac{\mu}{m_\chi^2}\left( \frac{\lambda}{2}-\frac{\mu^2}{m_\chi^2}  \right),\\
\mathcal{M}_{\chi\chi^{\dagger}\chi^{\dagger}\rightarrow\chi\chi} &=\frac{\mu}{m_\chi^2}\left( \frac{13}{24}\lambda-\frac{7\mu^2}{6m_\chi^2}  \right).
\end{align}
Assuming there is no CP violation, the conjugate processes will also have the same amplitude. The thermally averaged cross section for the $3\rightarrow 2$ process can be calculated by assuming Maxwell-Boltzmann distribution for the DM particles~\cite{Kuflik:2017iqs},
\begin{align}
\langle \sigma v^2 \rangle_{3\rightarrow 2}= \frac{\sqrt{5}}{2304\pi m_\chi^3}|\mathcal{M}|_{3\rightarrow 2}^2,
\end{align}
where the involved DM particles are taken to be non-relativistic.
The evolution of the $\chi$ and $\chi^\dagger$ number density can be obtained from the respective coupled Boltzmann equations. For the above-mentioned processes, relevant set of these equations can be given as follows,
\begin{align}
\frac{dN_\chi}{dt}=&\frac{1}{V_{\text{th}}^2}\bigg[-2\langle \sigma v^2 \rangle_{\chi\chi\chi\rightarrow\chi\chi^{\dagger}}\left( N_\chi^3 - n_\chi^{\text{eq}} N_{\chi}N_{\chi^\dagger} \right) 
+ \langle \sigma v^2 \rangle_{\chi^{\dagger}\chi^{\dagger}\chi^{\dagger}\rightarrow\chi\chi^{\dagger}}\left( N_{\chi^\dagger}^3 - n_\chi^{\text{eq}} N_{\chi}N_{\chi^\dagger}\right)\nonumber \\
&-2\langle \sigma v^2 \rangle_{\chi\chi\chi^\dagger\rightarrow\chi^{\dagger}\chi^{\dagger}}\left( N_\chi^2N_{\chi^\dagger} - n_\chi^{\text{eq}} N_{\chi^\dagger}^2 \right) 
+ \langle \sigma v^2 \rangle_{\chi\chi^\dagger\chi^\dagger\rightarrow\chi\chi}\left( N_\chi N_{\chi^\dagger}^2 - n_\chi^{\text{eq}} N_{\chi}^2 \right)\bigg]
\label{eq:Boltzmann_cann_particle},\\
\frac{dN_{\chi^\dagger}}{dt}=&\frac{1}{V_{\text{th}}^2}\bigg[\langle \sigma v^2 \rangle_{\chi\chi\chi\rightarrow\chi\chi^{\dagger}}\left( N_\chi^3 - n_\chi^{\text{eq}} N_{\chi}N_{\chi^\dagger} \right) 
-2 \langle \sigma v^2 \rangle_{\chi^{\dagger}\chi^{\dagger}\chi^{\dagger}\rightarrow\chi\chi^{\dagger}}\left( N_{\chi^\dagger}^3 - n_\chi^{\text{eq}} N_{\chi}N_{\chi^\dagger}\right) \nonumber \\
&+\langle \sigma v^2 \rangle_{\chi\chi\chi^\dagger \rightarrow \chi^{\dagger}\chi^{\dagger}}\left( N_\chi^2N_{\chi^\dagger} - n_\chi^{\text{eq}} N_{\chi^\dagger}^2 \right) 
-2 \langle \sigma v^2 \rangle_{\chi\chi^\dagger\chi^\dagger\rightarrow\chi\chi}\left( N_\chi N_{\chi^\dagger}^2 - n_\chi^{\text{eq}} N_{\chi}^2 \right) \bigg],
\label{eq:Boltzmann_cann_antiparticle}
\end{align}
where, $V_{\text{th}}$ is the volume of the core composed of thermalized DM particles inside the NS.

At this point we also note that the relevant $2\rightarrow2$ interaction of DM includes capture as well as annihilation into SM particles. The capture of ADM inside the NS can involve two mechanisms: (i) capture by elastic $\chi \text{SM}\rightarrow \chi \text{SM}$ scattering, where SM primarily include neutrons, and (ii) capture by DM self-scattering $\chi\chi \rightarrow \chi\chi$. The capture process by the scattering of DM with neutron is discussed in \cite{McDermott:2011jp, Bramante:2013hn, Bell:2013xk, Garani:2018kkd}, whereas the DM self-capture has been extensively analyzed in \cite{Guver:2012ba}. In this study, we consider the contribution of both of these mechanisms in the DM capture.

%%%%%%%%%%%%%%%%%%%%%%%%%%%%%%%%%
\subsection{Boltzmann equation for captured DM number}
%%%%%%%%%%%%%%%%%%%%%%%%%%%%%%%%%
The DM particles enter the NS surface with relatively high energy compared to the neutrons which are in the NS. Eventually, the DM particles loose momentum while undergoing scatterings with the neutrons and thermalize with the same kinetic energy. After thermalization, the DM particles tend to move towards the NS core and accumulate there. Due to this increasing density around the core, the DM-SM annihilation and $3\rightarrow2$ interactions take over and deplete the number. Depending on the interaction strength, the cannibal and annihilation rate can become equal to the rate of DM capture -- indicating an equilibrium and the total DM number attains a constant value. Here we introduce a quantity $f_\chi$, such that 
\begin{equation}
f_\chi=\frac{N_{\chi,\text{halo}}}{N_{\chi,\text{halo}}+N_{\chi^\dagger,\text{halo}}}~,
\end{equation}
which represents the relative fraction of DM particle in present in the DM halo within the galaxy. So, the initial particle-antiparticle asymmetry can quantified by defining $f_{\chi-{\chi^\dagger}}\equiv f_{\chi}-f_{\chi^\dagger}$. We will see that, this relative DM fraction and the asymmetry parameter will change as the DM accumulates inside the NS. In all our following discussions the $f_\chi$ parameter values represent only the initial DM particle abundance, which is kept as a free parameter. The complete evolution of DM numbers inside the NS is governed by the Boltzmann equations, which after taking into account all the $2\rightarrow2$ and $3\rightarrow2$ processes, can be given as,
\begin{subequations}
\label{eq:Boltzmann_eq}	
\begin{align}
\frac{dN_\chi}{dt} &= f_\chi C_B + C_{\chi\chi} N_\chi - \frac{\langle \sigma v \rangle_\text{ann}}{V_\text{th}}N_{\chi}N_{\chi^\dagger} + \Bigg( \frac{dN_\chi}{dt} \Bigg)_{3\rightarrow2}, \\
\frac{dN_{\chi^\dagger}}{dt} &= (1-f_\chi) C_B + C_{\chi\chi} N_{\chi^\dagger} - \frac{\langle \sigma v \rangle_\text{ann}}{V_\text{th}}N_{\chi}N_{\chi^\dagger} + \Bigg( \frac{dN_{\chi^\dagger}}{dt} \Bigg)_{3\rightarrow2},
\end{align}
\end{subequations}
where $C_B$ and $C_{\chi\chi}$ are the capture rates by DM-neutron scattering and DM self-scattering respectively, and $\langle \sigma v \rangle_\text{ann}$ incorporates all kinematically feasible annihilation channels. The last terms in both of these equations, i.e., $\big( dN_{\chi,\chi^\dagger}/dt \big)_{3\rightarrow2}$ are shown in Eqs. \eqref{eq:Boltzmann_cann_particle} and \eqref{eq:Boltzmann_cann_antiparticle}. 
Treating DM as a non-relativistic scalar the capture rate $C_B$ due to $\chi n\rightarrow \chi n$ elastic scattering can be expressed as \cite{McDermott:2011jp},
\begin{equation}
C_B=\sqrt{\frac{6}{\pi}}\frac{\rho_\text{DM}}{m_\chi}\frac{v_\text{esc}^2}{\bar{v}^2}\big( \bar{v}\sigma_{\chi n} \big)\xi  N_B \left( 1-\frac{1-\exp(-B^2)}{B^2}  \right),
\label{eq:DM_Capture_rate}
\end{equation}
where $\sigma_{\chi n}$ is the elastic scattering cross-section between DM and nuclear matter inside the NS. We take some generic values for the escape velocity at the NS surface $v_\text{esc}=1.8\times10^5$ km/s and average velocity of DM particles within the halo $\bar{v}=220$ km/s~\cite{McDermott:2011jp}. $N_B (\sim 1.7\times10^{57})$ is the total number of baryonic particles within the NS. The factor $B$ is defined as
%
%\begin{equation}
$B^2=\frac{3}{2}\frac{v(r)^2}{\bar{v}^2}\frac{\mu}{\mu_-^2}$,
%\end{equation}
%
where $\mu=m_\chi/m_n$ and $\mu_-=(\mu-1)/2$, $m_n$ being the mass of neutron. Denoting the momentum transfer due to $\chi-n$ scattering by $\delta p$, the quantity $\xi$ can be written as,
%
%\begin{equation}
$\xi= \min[1,\delta p/p_F]$
%\end{equation}
%
where $p_F$ is the Fermi momentum corresponding to the standard nuclear matter within the NS given by $(3\pi^2\rho_B/m_n)^{1/3}\simeq0.575$ GeV. The baryonic matter density inside NS, taken to be $\rho_B=1.5\times10^{14}$ g/cm$^3$. The momentum transfer can be expressed as $\delta p\simeq \sqrt{2}m_r v_\text{esc}$, where $m_r$ is the reduced mass of $m_\chi$ and $m_n$. The total DM density within the MW galaxy is taken to be $\rho_\text{DM}\simeq 1$ GeV/cm$^3$. Recently, an improved treatment of capture rate $C_B$, taking account of the Pauli blocking effect, and the enhancement due to bosonic DM is presented in~\cite{Garani:2018kkd}. In our case, we aim to investigate the implications of cannibal dark matter of masses around 1 GeV, where a classical Maxwell-Boltzmann treatment is sufficient. 
The process of DM self-capture by $\chi\chi\rightarrow\chi\chi$ scatterings can also enhance the number of captured DM significantly. This type of capture rate is given by \cite{Guver:2012ba},
\begin{equation}
C_{\chi\chi}=\sqrt{\frac{3}{2}}\frac{\rho_\text{DM}}{m_\chi}\sigma_{\chi\chi}v_\text{esc}(R)\frac{v_\text{esc}(R)}{\bar{v}}\langle \hat{\phi}_\chi \rangle \frac{\erf{(\eta)}}{\eta}\frac{1}{1-\frac{2GM}{R}}
\end{equation}
where $\sigma_{\chi\chi}$ is the DM self-scattering cross section. The quantity $\langle \hat{\phi}_\chi \rangle = v_{\text{esc}}^2(r)/v_{\text{esc}}^2(R)$ represents the compactness of the star, which as a conservative choice can be taken to as unity, assuming the baryonic matter density to be uniform inside the star. The quantity $\eta^2\equiv 3/2(v_N/\bar{v})^2$, with $v_N$ being the velocity of NS in the galaxy. The self-capture of DM continues until the rate saturates the geometrical limit, 
%
%\begin{equation}
$\left(N_\chi\sigma_{\chi\chi}\right)_\text{max}=\pi r_{\chi}^2$,
%\end{equation}
% 
where $r_{\chi}$ is the radius of the DM core inside the NS. For the interactions shown in the Lagrangian~\eqref{eq:Lagrangian}, DM self-scattering cross section can be evaluated as,
\begin{equation}
\sigma_{\chi\chi}=\frac{1}{64\pi^{3/2}m_\chi^2}\Bigg( \lambda - \frac{5\mu^2}{3m_\chi^2} \Bigg)^2. 
\end{equation}
After entering the neutron star (NS), dark matter (DM) particles lose energy through repeated $\chi- n$ scatterings and eventually thermalize with the baryonic matter in the stellar interior. This thermalization process has been extensively studied in Ref.~\cite{Bertoni:2013bsa} and more recently in Ref.~\cite{Garani:2018kkd}. This effectively is a two-step process: (i) following the initial collision with the NS, DM particles undergo multiple revolutions around the star, gradually losing energy with each subsequent scattering until they become gravitationally bound and trapped within the NS; and (ii) once trapped, the DM particles further thermalize via continued collisions with dense nuclear matter. This framework was further extended in Ref.~\cite{Garani:2020wge}, incorporating a range of DM candidates and various mediators for DM-SM interactions inside the NS.
\begin{figure}[htbp]
\centering
\includegraphics[scale=0.2]{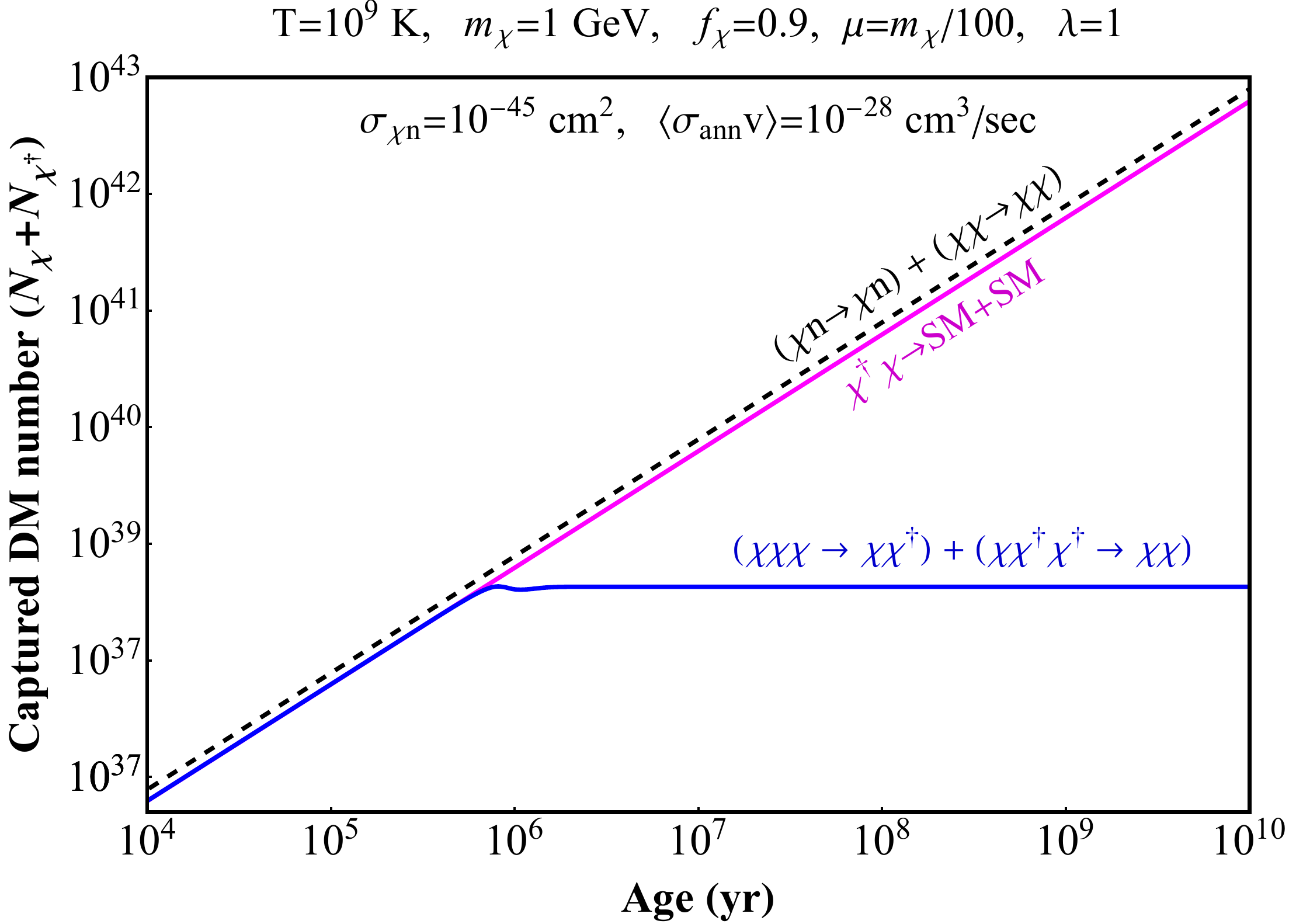}~~
\includegraphics[scale=0.2]{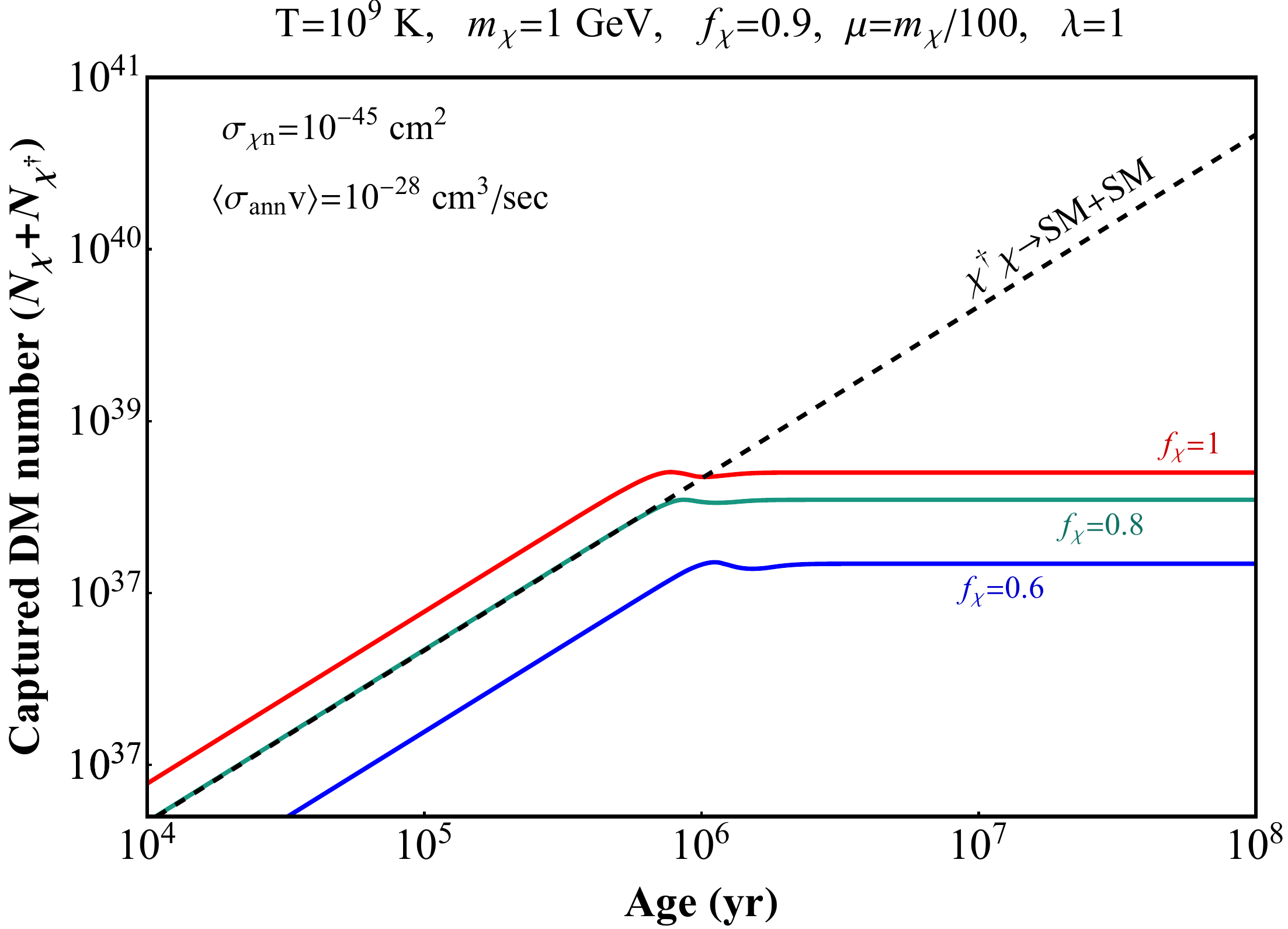}
\caption{\small{\textbf{(Left)} Evolution of total DM number ($N_{\chi}+N_{\chi^\dagger}$) inside the neutron star ($N_\chi$) has been plotted for different DM interactions. The blue and magenta solid lines indicate DM number for $3\rightarrow2$ and DM-SM annihilation, respectively. The black dashed line indicates the DM number due to DM capture only. Here, the DM mass $m_\chi$ has taken to be 1 GeV, DM quartic self-coupling $\lambda=1$ and the interior temperature of NS is $T=10^9$ K. The initial relative DM particle fraction in halo is taken to be $f_\chi=0.9$. The DM-SM scattering and DM annihilation cross sections are taken to be $\sigma_{\chi n}=10^{-45}~\text{cm}^2$ and $\sigma_\text{ann}=10^{-28}~\text{cm}^2$.  \textbf{(Right)} Evolution of total DM number with the variation of initial DM particle relative abundance ($f_\chi$) are plotted. The blue, green, and red lines indicate the DM number for $f_\chi=0.6,~0.8~\text{and}~1$ respectively. The black dashed line shows the DM number evolution by taking only the DM-SM annihilation, for comparison. All the relevant parameters are the same as the left panel.}}
\label{fig:Evolution_DM_Number}
\end{figure}
In this work, we adopt the framework developed in Ref.~\cite{Garani:2018kkd} to compute the thermalization time and to determine the radius of the DM core formed by the thermalized particles. The thermalization time can be approximated as~\cite{Garani:2018kkd},
\begin{equation}
t_\text{th}\simeq 1.07\times 10^4 \frac{\gamma}{(1+\gamma)^2}\Bigg( \frac{10^5~\rm K}{T} \Bigg)^2 \Bigg( \frac{10^{-45}~\rm cm^2}{\sigma_{\chi n}}  \Bigg)~~\rm yrs,
\end{equation}
where $\gamma=m_\chi/m_n$. After thermalization, the captured dark matter migrates toward the stellar interior and accumulates to form a dense DM core. The thermalization timescale becomes particularly relevant at late stages of evolution, when the NS core temperature drops to $\mathcal{O}(10^3)\,\mathrm{K}$. For the benchmark parameters considered here, the thermalization time at a core temperature of $T = 2500\,\mathrm{K}$ is $t_{\rm th} = 1.7 \times 10^7\,\mathrm{yr}$, which is approximately an order of magnitude smaller than the characteristic evolutionary timescale ($\sim 10^8\,\mathrm{yr}$). Therefore, the assumption of efficient thermalization remains valid in this regime. However, for smaller DM–nucleon scattering cross sections, the thermalization process can be significantly delayed, and a more careful treatment of the non-instantaneous thermalization of DM particles is required.
%
%\newline
The characteristic radius of this thermalized DM core is given by~\cite{Garani:2018kkd},
\begin{equation}
R_\text{th} = \sqrt{\frac{9T}{4\pi G\rho_B m_\chi}}= 4.29\left( \frac{T}{10^5 \rm~K} \right)^{1/2}\left( \frac{1\rm~ GeV}{m_\chi}\right)^{1/2}~~\rm{meter}.
\end{equation}
This radius has been used to compute the volume of thermal DM core in Eq.~\eqref{eq:Boltzmann_eq}. We show the evolution of captured DM number inside a NS core by taking into account both the $\chi\chi\rightarrow\chi\chi$ self-scattering and $\chi n\rightarrow\chi n$ scattering in Fig.~\ref{fig:Evolution_DM_Number}. The DM-SM elastic scattering and DM annihilation to SM cross sections are chosen to be $\sigma_{\chi n}=10^{-45}~\text{cm}^2$ and $\sigma_\text{ann}=10^{-28}~\text{cm}^2$ which are well within the current direct and indirect detection bounds \cite{Andrade:2020lqq, Eckert:2022qia, XENON:2023cxc, HAWC:2017mfa, Hess:2021cdp, Fuke:2005it, Arguelles:2019ouk, DAmico:2018sxd, LZ:2022lsv, Billard:2021uyg}. We take the internal temperature of the NS to be $T=10^9$ K. The DM mass is taken to be $m_\chi=$ 1 GeV where the quantum corrections for the capture rates become  less significant~\cite{Garani:2018kkd}. The total dark matter (DM) number increases due to capture via DM-neutron scattering and DM self-scattering, while it decreases through $3\to 2$ cannibalization and DM annihilation. An equilibrium is eventually established in this interplay, depending on the strength of various DM interaction couplings. 
We emphasize that the evolution of the DM number shown in Fig.~\ref{fig:Evolution_DM_Number} depends explicitly on the thermal volume of the DM core through Eq.~\eqref{eq:Boltzmann_eq}, which is itself a temperature-dependent quantity. In the present section, however, we have not modeled the thermal evolution of the NS core. To maintain a realistic description of the captured DM population, we therefore adopt an externally motivated NS temperature profile taken from Fig.~\ref{fig:NS_Temperature_evolution}, fixing the late-time temperature to be $ 4000~\mathrm{K}$. In the following section, we incorporate the thermal evolution of the NS self-consistently by solving the coupled differential equations governing both the NS temperature and the DM number. 
From the left plot in Fig.~\ref{fig:Evolution_DM_Number}, we find that for the chosen DM-SM cross sections and self-coupling parameters, cannibal reaction plays a dominant role in determining the final DM abundance. The right plot of Fig.~\ref{fig:Evolution_DM_Number} shows the evolution of the DM number inside the core for different values of the initial retive abundance of DM particle $f_\chi$, keeping the DM couplings fixed. As the asymmetry increases, thereby increasing the effective DM capture rate (i.e. $f_\chi C_B$), which results in higher accumulated DM number. The blue, green and red lines indicate the DM number for $f_\chi=0.6,~0.8,~\text{and}~1$ respectively. The black dashed line represents the DM number due to annihilation channel only.
\begin{figure}[t]
\centering
\includegraphics[scale=0.3]{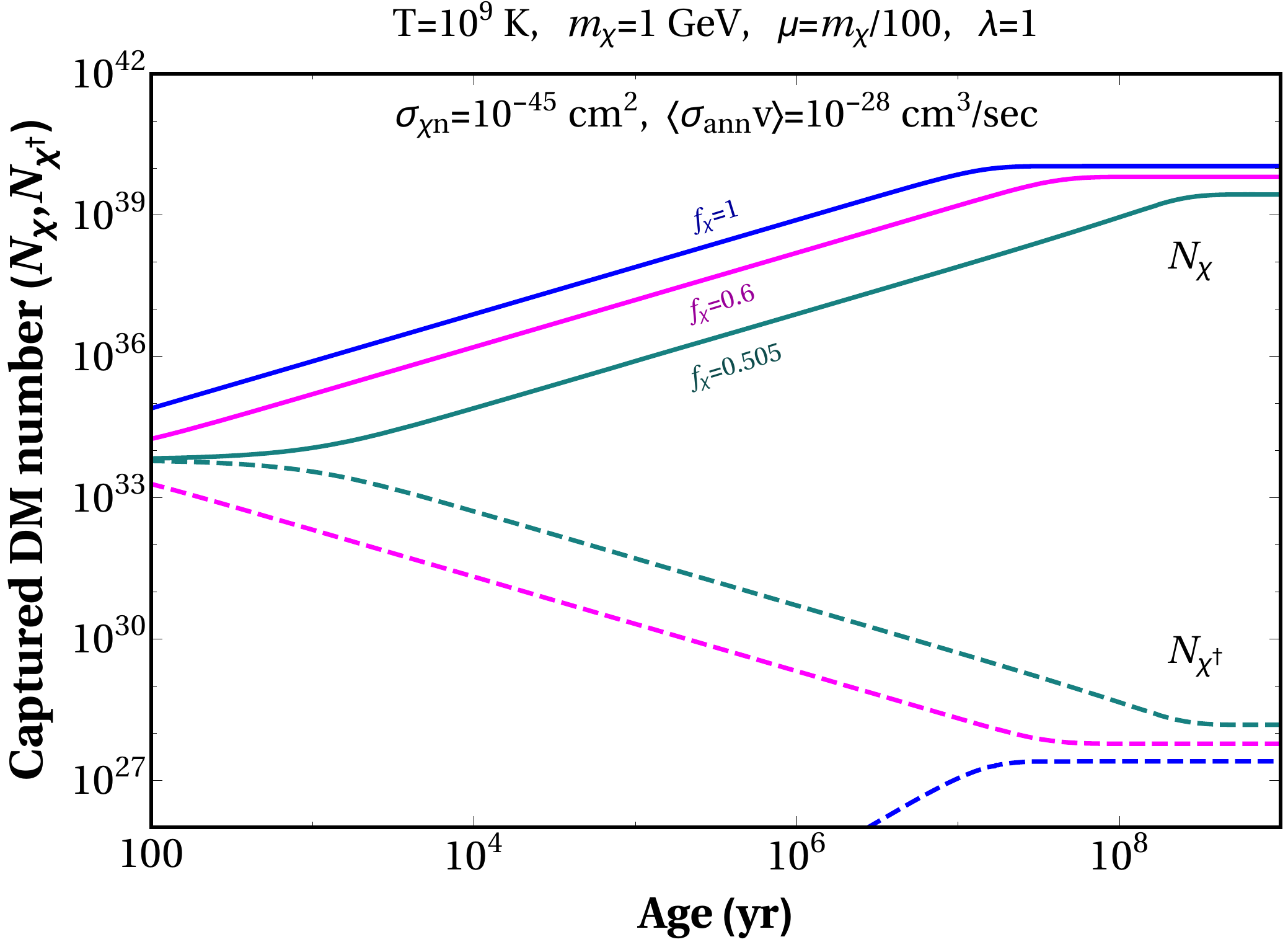}
\caption{\small{Evolution of the number of DM particle and antiparticle is plotted for different initial DM particle fractions $f_\chi$. The green, magenta and blue lines are plotted with $f_\chi=0.505,~0.6$ and 1 respectively. The solid and dashed lines represent the number of particle ($N_\chi$) and antiparticle ($N_{\chi^\dagger}$) respectively. All the relevant parameters are same as Fig.~\ref{fig:Evolution_DM_Number}. }}
\label{fig:DM_Number_asymmetry}
\end{figure}

As shown in Eqs.~\eqref{eq:Boltzmann_cann_particle} and \eqref{eq:Boltzmann_cann_antiparticle}, both particle and antiparticle abundances play essential role in determining the evolution of the total DM population inside the NS. To illustrate this, we plot in Fig.~\ref{fig:DM_Number_asymmetry} the time evolution of $N_{\chi}$ and $N_{\chi^\dagger}$ for different initial relative particle abundances. The green, magenta, and blue curves correspond to $f_\chi = 0.505,~0.6$, and $1$, respectively, with solid and dashed lines denoting $N_\chi$ and $N_{\chi^\dagger}$. Even for a small initial relative particle fraction ($f_\chi = 0.505$), $\chi^\dagger \chi \rightarrow \mathrm{SM}+\mathrm{SM}$ annihilations rapidly deplete the antiparticle abundance, thereby enhancing the asymmetry during the evolution. Furthermore, we observe that even in a fully asymmetric scenario, i.e., $(N_{\chi^\dagger})_{\mathrm{initial}} = 0$, the $3 \rightarrow 2$ process can regenerate a nonzero antiparticle population.

%%%%%%%%%%%%%%%%%%%%%%%%%%%%%%
\section{Evolution of Neutron Star temperature} 
\label{sec:Evolution_NS_temperature}
%%%%%%%%%%%%%%%%%%%%%%%%%%%%%%

%%%%%%%%%%%%%%%%%%%%%%%%%%%%%%
\subsection{Heating by DM}
\label{subsec:NS_heating_DM}
%%%%%%%%%%%%%%%%%%%%%%%%%%%%%%
Although DM constitutes only a small fraction of a neutron star's mass, it can nevertheless heat the star significantly, thereby modifying its present‐day surface temperature. In the absence of annihilation into SM particles, ADM heats the neutron star predominantly via three mechanisms: (i) Kinetic heating: the captured dark matter particles, being more energetic than the neutrons, shed their kinetic energy through elastic scattering with the NS matter \cite{Baryakhtar:2017dbj,Zhitnitsky:2023fhs,Acevedo:2019agu,Raj:2017wrv}. (ii) Annihilation heating: the DM can annihilate to SM particles \cite{Bell:2023ysh}. These SM particles can get thermalized with the stellar constituents to rise the interior temperature of NS. (iii) Cannibal heating: the $3\rightarrow2$ processes produce final state particles with elevated energy. These high energy particles eventually transfer the residual energy to the SM particles by elastic scattering, thereby increasing the NS temperature. In this study we focus on the post‑thermalization evolution of the DM population within the NS and, accordingly, explore the possibility where cannibalization can be the dominant mechanism governing the star’s DM–induced heating.
The energy generated per unit time due to the $3\to2$ processes can be given as,
\begin{align}
\big(\Delta E\big)_{3\rightarrow2} & =  m_\chi N_\chi \Gamma_{3\rightarrow2} \nonumber \\
& = \frac{m_\chi}{V_\text{th}^2} \left[\langle \sigma v^2 \rangle_{\chi\chi\chi\rightarrow\chi\chi^{\dagger}}\Big( N_{\chi}^3 + N_{\chi^\dagger}^3 \Big) + \langle \sigma v^2 \rangle_{\chi\chi^{\dagger}\chi^{\dagger}\rightarrow\chi\chi}\Big( N_{\chi}^2N_{\chi^\dagger} + N_{\chi}N_{\chi^\dagger}^2 \Big)\right],
\label{eq:emissivity-DM}
\end{align}
were $\Gamma_{3\rightarrow2}$ is the cannibal reaction rate. Here we have ignored the back reactions from Eqs.~\eqref{eq:Boltzmann_cann_particle} and \eqref{eq:Boltzmann_cann_antiparticle} because of the exponential suppression of $n_{\chi}^{\text{eq}}$. The DM–induced energy emissivity is defined as the amount of energy produced per unit time per unit volume within the NS, 
\begin{equation}
\epsilon_{\text{DM},3\rightarrow2}=\frac{\big(\Delta E\big)_{3\rightarrow2}}{(4/3)\pi R^3},
\label{eq:emissivity_DM_cann}
\end{equation}
where $R$ is the NS radius. The emissivity due to DM annihilation into the SM particles is given by,
\begin{equation}
\epsilon_\text{DM,ann}=\frac{2m_\chi}{(4/3)\pi R^3}\frac{\langle \sigma v \rangle_\text{ann}}{V_{\text{th}}}N_{\chi}N_{\chi^\dagger},
\label{eq:emissivity_DM_ann}
\end{equation}
Here we have taken the annihilation cross section to be $\langle \sigma v \rangle_\text{ann}=10^{-28}~\text{cm}^2$ throughout the paper. Kinetic heating arises during the phase in which DM particles are captured and subsequently lose energy through elastic scattering with neutrons inside the NS. This mechanism deposits energy into the stellar interior until the DM population becomes thermalized. The corresponding rate of energy deposition due to DM-neutron elastic scattering is given by $\mathcal{K}_{\rm DM} = C_B \langle E_R \rangle$ where $C_B$ denotes the DM capture rate via $\chi$--$n$ elastic scattering, as defined in Eq.~\eqref{eq:DM_Capture_rate}, and $\langle E_R \rangle$ represents the average recoil energy transferred from the DM particle to the neutron per collision. The expression for $\langle E_R \rangle$ can be written as \cite{Keung:2020teb},
\begin{equation}
\langle E_R \rangle \equiv \frac{\int_{-1}^{1}d\cos\theta_{\text{cm}}E_R\frac{d\sigma_{\chi n}}{d\cos\theta_{\text{cm}}}}{d\cos\theta_{\text{cm}}\frac{d\sigma_{\chi n}}{d\cos\theta_{\text{cm}}}} \simeq 
\frac{(1-\bar{B})m_\chi \mu}{\bar{B}+2\sqrt{\bar{B}}\mu+\bar{B} \mu^2}
\end{equation}
where, $\bar{B}=1-2GM/(c^2R)$ and $\mu=m_\chi/m_n$. The emissivity due to kinetic heating is given by,
\begin{equation}
\epsilon_\text{DM,kin}= \frac{C_B \langle E_R \rangle}{(4/3)\pi R^3}.
\label{eq:emissivity_DM_kin}
\end{equation}
\begin{figure}[t]
\centering
\includegraphics[scale=0.35]{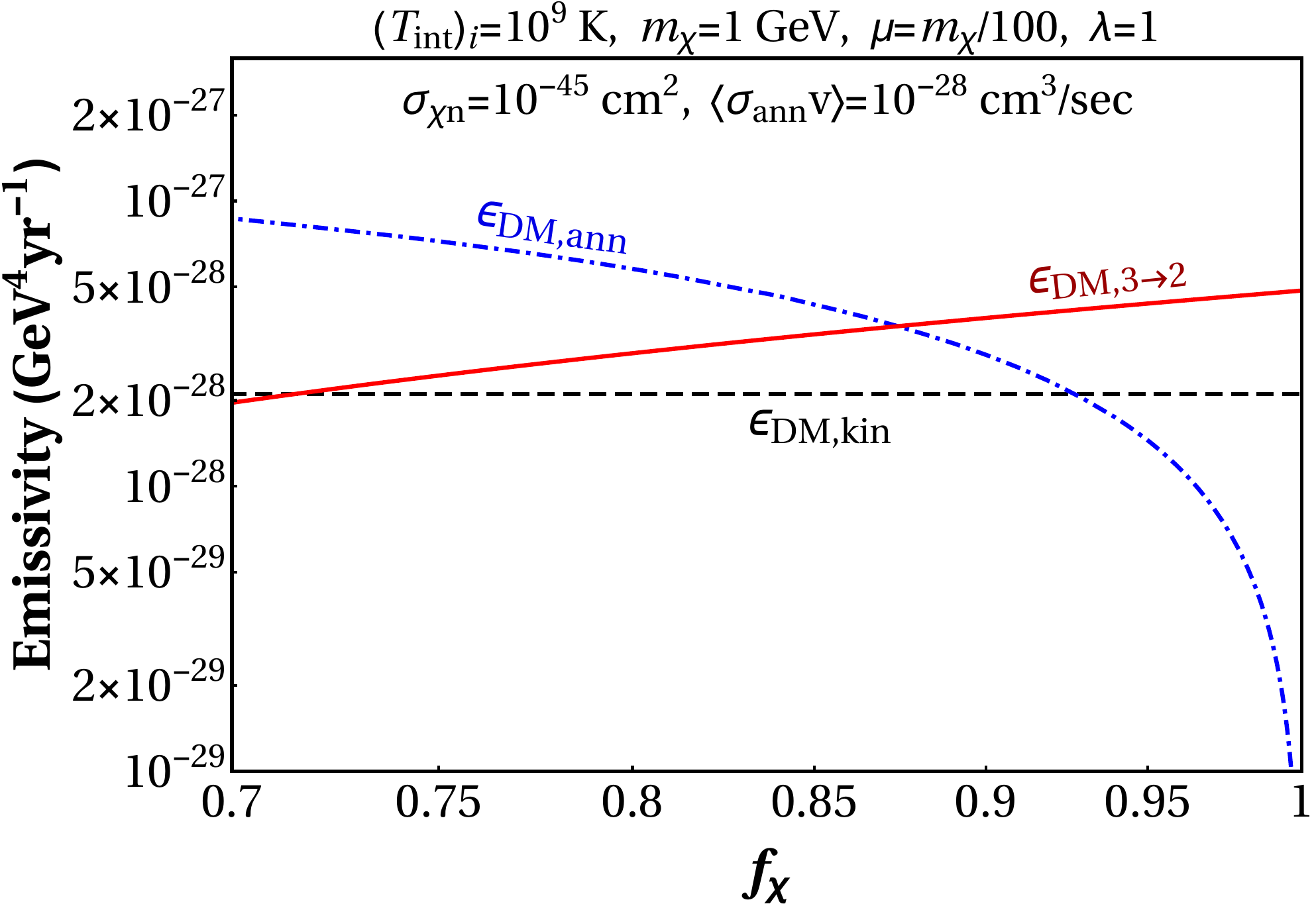}~
\caption{\small{ The thermal emissivity arising from DM kinetic and cannibal heating, as a function of initial DM particle relative abundance~($f_\chi$) are plotted. The red solid and blue dot-dashed lines correspond to the emissivity for cannibal and annihilation heating, respectively. The kinetic heating emissivity is shown by black dashed line for reference. The initial NS interior temperature is taken to be $10^9$ K and the DM quartic and trilinear coupling $\lambda=1,\mu=m_\chi/100$  and mass $m_\chi$=1 GeV. The DM cross-sections are same as Fig.~\ref{fig:Evolution_DM_Number}.}}
\label{fig:Emissivity_kinetic_cannibal}
\end{figure}
In order to identify the regime where either annihilation or cannibal heating dominates the NS temperature evolution, we present in Fig.~\ref{fig:Emissivity_kinetic_cannibal} the emissivities evaluated at $t = 10^9~\mathrm{yr}$. The cannibal emissivity, $\epsilon_{\mathrm{DM},\,3\rightarrow 2}$, is shown as a function of the initial DM particle relative fraction $f_\chi$ for $m_\chi = 1~\mathrm{GeV}$, and compared with the annihilation emissivity, $\epsilon_{\mathrm{DM},\,\mathrm{ann}}$. As the asymmetry increases, the number of accumulated DM particles rises due to the enhanced effective capture rate, as illustrated in the right panel of Fig.~\ref{fig:Evolution_DM_Number}. Consequently, the emissivity associated with the $3\rightarrow 2$ process grows for a fixed cross section. In contrast, the increasing asymmetry suppresses the $\chi^\dagger$ abundance, thereby reducing the probability of $\chi$--$\chi^\dagger$ annihilation. This leads to a rapid decrease in the annihilation emissivity. We find that cannibal heating becomes the dominant contribution for $f_\chi \gtrsim 0.85$.

%

%

%%%%%%%%%%%%%%%%%%%%%%%%%%%%%%
\subsection{Cooling by photon  and neutrino} 
\label{sec:NS_cooling_photon_neutrino}
%%%%%%%%%%%%%%%%%%%%%%%%%%%%%%
Since the formation of a NS, energy is being carried away mainly by photons and neutrinos. For the first million years (Myr) after formation, the neutrinos are produced from very dense nuclear matter by direct Urca process. The energy emitted per unit volume, i.e. the emissivity due to this process varies as, $\epsilon_\nu\sim T^6$. For the less dense NS matter, the energy loss is primarily taken over by modified Urca process where the temperature varies as $\epsilon_\nu\sim T^8$. The emissivity can be written as \cite{Shapiro:1983du,Kouvaris:2007ay,Keung:2020teb},
\begin{equation}
\epsilon_\nu \simeq 1.81\times 10^{-27} ~ \bigg(\frac{n_F}{n_0}\bigg)^{2/3}\bigg( \frac{T_{\text{int}}}{10^7~K} \bigg)^8 ~\rm GeV^4yr^{-1}
\label{eq:emissivity-Neutrino}
\end{equation}
where $n_0=0.16\times10^{39}~\rm cm^{-3}$ and $n_F$ is the number density of nuclear matter inside the NS. Here we denote the temperature of neutron star's interior by $T_{\text{int}}$. 
Once the stellar temperature falls below $10^8$ K -- typically after the first Myr, the cooling is governed predominantly by photon emission from the surface of NS rather than by neutrino losses. The photon emissivity can be expressed in terms of the interior temperature of NS as \cite{Keung:2020teb},
\begin{equation}
\epsilon_\gamma \simeq \begin{cases}
2.59 \times10^{-17}\bigg( \frac{T_{\text{int}}}{10^8K} \bigg)^{2.2}~\text{GeV}^4~\text{yr}^{-1} ,~~T_{\text{int}}\geq3700~K, \\
2.44 \times10^{-9} \bigg( \frac{T_{\text{int}}}{10^8K} \bigg)^{4}~\text{GeV}^4~\text{yr}^{-1} ,~~~~~T_{\text{int}}<3700~K. \end{cases}
\label{eq:emissivity-Photon}
\end{equation}
Present day observational constraints on NS are set by their measured surface temperatures. Since the surface temperature is governed by the thermal state of the stellar interior, we proceed to model the evolution of interior temperature ($T_{\text{int}}$), incorporating both the standard cooling processes and the heating produced by DM-induced energy deposition,
\begin{equation}
\label{eq:NS_Temp_Evolution}
\frac{dT_\text{int}}{dt}=\frac{1}{c_V}(-\epsilon_\nu -\epsilon_\gamma + \epsilon_{\text{DM}}),
\end{equation}
where $\epsilon_{\nu,\gamma}$ represents the neutrino and photon emissivities described in Eqs.~\eqref{eq:emissivity-Neutrino} and \eqref{eq:emissivity-Photon} respectively. $\epsilon_{\text{DM}}$ incorporates the total emissivity due to heating caused by interactions involving DM : $\epsilon_{\text{DM}}=\epsilon_{\text{DM,kin}}+\epsilon_{\text{DM,ann}}+\epsilon_{\text{DM},3\rightarrow2}$ mentioned in Eqs.~\eqref{eq:emissivity_DM_kin}, \eqref{eq:emissivity_DM_ann} and \eqref{eq:emissivity_DM_cann}, respectively. Considering the neutrons as ideal Fermi gas, the specific heat $c_V$ can be given by $c_V= (k_B^2T_\text{int}/3)p_{F,n}\sqrt{m_n^2+p_{F,n}^2}$ with the Fermi momenta for neutron $p_{F,n}=0.34\left(n_F(1-a_\chi)/n_0  \right)^{1/3}~\rm GeV$ \cite{Kouvaris:2007ay, Keung:2020teb}.
Here, $a_\chi$ denotes the fraction of DM particles relative to the total number of constituents of the NS matter, $a_\chi=N_\chi/(N_{\chi}+N_{\text{SM}})$ and $N_{\text{SM}}$ is the total number of SM fermions present in the stellar interior. Usually, astrophysical observations concern the surface temperature ($T_{\text{sur}}$), which is related to $T_{\text{int}}$ as \cite{Keung:2020teb},
\begin{equation}
\label{eq:T_surface}
T_{\text{sur}}= \begin{cases}
0.87\times 10^6~\text{K} \left( \frac{g_s}{10^{14}~\text{cm} ~ \text{s}^{-2}} \right)^{1/4}\left( \frac{T_{\text{int}}}{10^8 \text{K}} \right)^{0.55},~~T_{\text{int}}\gtrsim 3700 ~ \text{K}, \\
T_{\text{int}}, ~~~~T_{\text{int}}\lesssim 3700 ~ \text{K},
\end{cases}
\end{equation}
where $g_s=GM/R^2$ is the gravitational acceleration at the surface of NS. The observed temperature of NS ($T_{\text{obs}}$), after accounting for gravitational redshift, is related to $T_{\text{sur}}$ via 
%
%\begin{equation}
%\label{eq:T_observed}
$T_{\text{obs}}=T_{\text{sur}}\sqrt{1-2GM/R}$.
%\end{equation}
%
% 
Note that Eq.~\eqref{eq:Boltzmann_eq} shows that the evolution of DM number depends on the stellar interior temperature, which enters through both the capture rate and the thermal radius of DM core within which the $3\rightarrow2$ reactions deplete the DM population. Conversely, the temperature evolution depends on the DM number via the cannibal reaction rate $\Gamma_{3\rightarrow2}$, as seen from Eq.~\eqref{eq:NS_Temp_Evolution}. Therefore, to obtain the full evolution of the DM abundance and the interior temperature, one must solve the coupled differential Eqs.~\eqref{eq:Boltzmann_eq} and~\eqref{eq:NS_Temp_Evolution}.
\begin{figure}[t]
\centering
\includegraphics[scale=0.25]{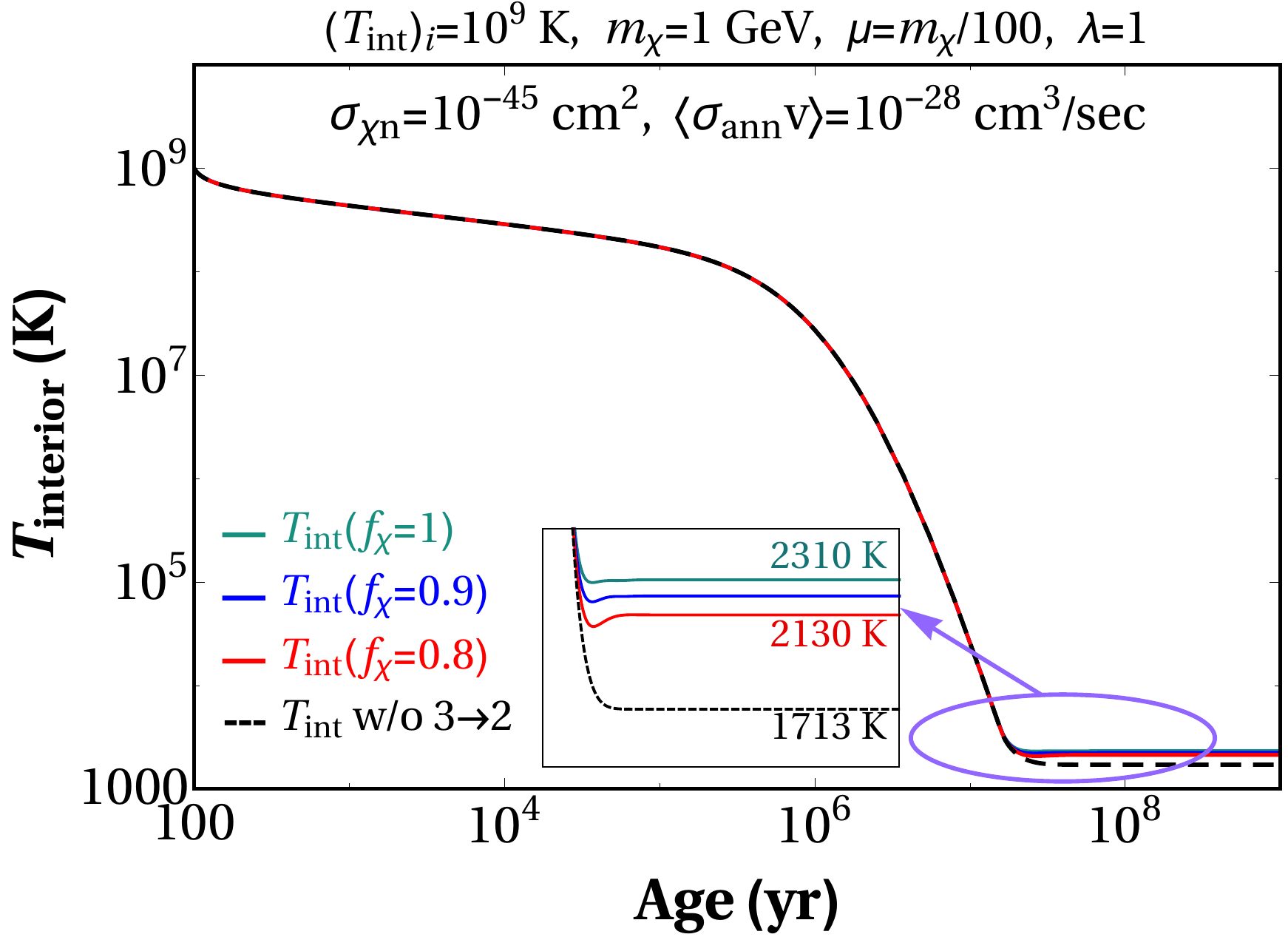}~~
\includegraphics[scale=0.24]{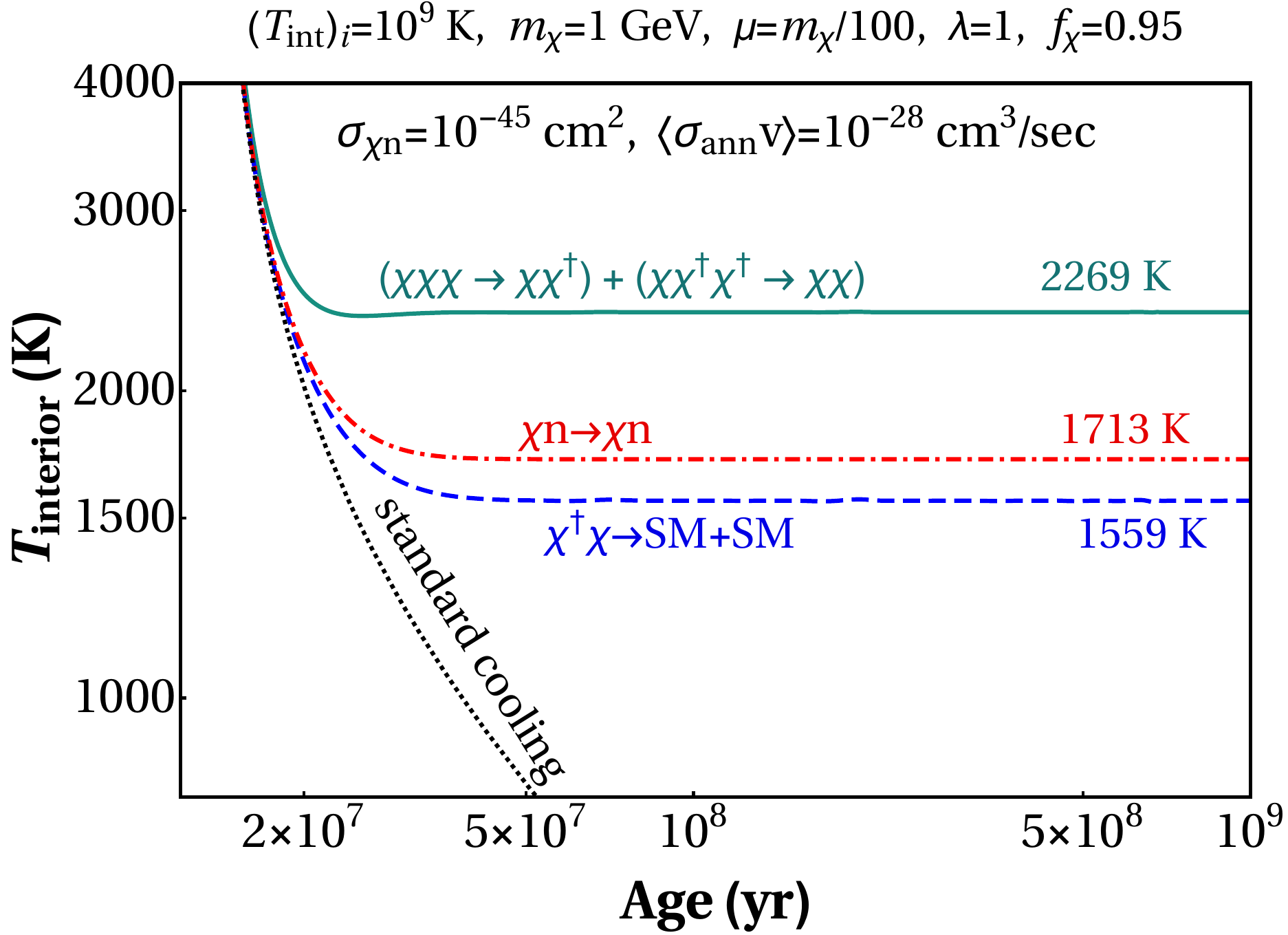}
\caption{\small{\textbf{(Left)} Evolution of NS interior temperature is shown with variations of initial DM particle fraction $f_\chi$. The black dashed line represents the NS temperature without the presence of $3\rightarrow2$ and DM annihilation reaction. The initial internal temperature is taken to be, $(T_{\text{int}})_i=10^9$ K and DM mass is 1 GeV. All the other parameters are the same as Fig.~\ref{fig:Evolution_DM_Number}.  \textbf{(Right)} Comparison between the evolution of $T_{\text{int}}$ for cannibal, annihilation, and kinetic heating, indicated by green, blue, and red lines, respectively. The resultant interior temperatures are also indicated for comparison. The black dotted line represents the temperature without any heating due to DM. See text for further details.}}
\label{fig:NS_Temperature_evolution}
\end{figure}
The full evolution of the neutron star (NS) interior temperature, $T_{\text{int}}$, is shown in the left panel of Fig.~\ref{fig:NS_Temperature_evolution}. For a NS with mass $1.4M_{\odot}$ and radius 10~km, the initial interior temperature is taken to be $10^9~\mathrm{K}$. The black dashed line represents the kinetic heating scenario, incorporating $\chi n\rightarrow\chi n$ scattering. To account for additional heating due to DM cannibal reactions, we consider a DM mass of 1~GeV, DM trilinear coupling $\mu=m_\chi/100$ and quartic coupling $\lambda=1$. As the emissivities due to neutrino and photon emission scale with temperature (see Eqs.~\eqref{eq:emissivity-Neutrino} and~\eqref{eq:emissivity-Photon}), these processes become inefficient as the NS cools. Subsequently, depending on the $3 \to 2$ reaction rate, the cannibal heating dominates, leading to a plateau in temperature. The red, blue, and green solid lines correspond to interior temperature profiles for initial DM particle fraction $f_\chi = 0.8, 0.9  ~\text{and}~ 1$, respectively, corresponding asymptotic values of $T_{\text{int}}$ are mentioned in the figure.

In the right panel of Fig.~\ref{fig:NS_Temperature_evolution}, we present the comparative contributions of various DM-induced heating mechanisms arising from scattering and annihilation processes. The evolution of the internal temperature $T_{\rm int}$ in the absence of any heating, corresponding to standard cooling, is indicated by the black dotted curve. Heating due to DM annihilation $\chi^\dagger \chi \rightarrow {\rm SM+SM}$ is subdominant for the initial relative $\chi$ abundance $f_\chi=0.95$, as shown by the red dot-dashed curve. The kinetic heating is represented by the blue dashed curve for DM-SM elactic cross section $\sigma_{\chi n}=10^{-45}~\text{cm}^2$. Finally, the green curve corresponds to heating from number-changing $3 \rightarrow 2$ processes. For the benchmark values of the DM couplings $\mu=m_\chi/100, ~\lambda=1$ and cross sections taken here, the dominant contribution to NS heating arises from the cannibal reaction, which ultimately governs the rise in temperature.

\begin{figure}[htbp]
	\centering
	\includegraphics[scale=0.3]{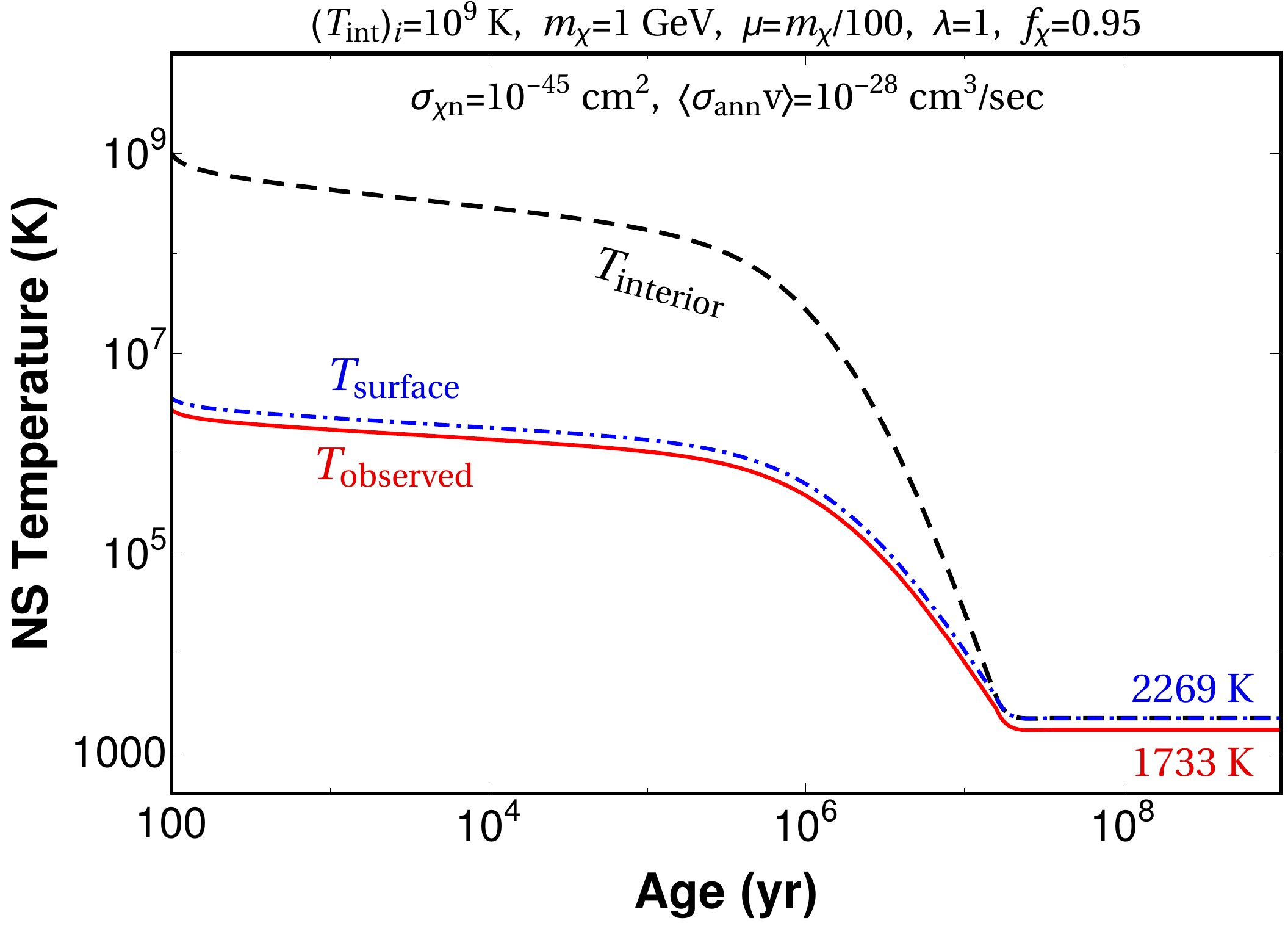}
	\caption{\small{Comparison between the interior, surface and observed temperature of NS. We fix the DM trilinear coupling to be $\mu/m_\chi=10^{-2}$, and DM mass $m_\chi=1$ GeV. All the other parameters are same as Fig. \ref{fig:NS_Temperature_evolution}.}}
	\label{fig:NS_Temperature_comparison}
\end{figure}

In Fig.~\ref{fig:NS_Temperature_comparison} we show the evolution of $T_{\text{int}}$, $T_{\text{sur}}$, and $T_{\text{obs}}$, indicated by the black dashed, blue dot-dashed, and red solid lines, respectively. For this plot, the DM mass is taken to be $m_\chi = 1~\mathrm{GeV}$, with an initial DM particle abundance $f_\chi = 0.95$. For these parameter choices, the interior temperature stabilizes at $T_{\text{int}} \approx 2269~\mathrm{K}$, which eventually matches the surface temperature, as defined in Eq.~\eqref{eq:T_surface}. Due to gravitational redshift, the observed temperature asymptotically settles to a slightly lower value, $T_{\text{obs}} \approx 1733~\mathrm{K}$.

%%%%%%%%%%%%%%%%%%%%%%%%%%%%%% 
\section{Prospects at JWST, ELT, and TMT} 
\label{sec:Observation_NS_temperature}
%%%%%%%%%%%%%%%%%%%%%%%%%%%%%% 
The surface temperature of neutron stars (NSs) decreases from approximately $10^6$~K in young stars (with ages less than $\sim 1$~Myr) to about $10^3$~K in older stars (ages of several Gyr) \cite{Potekhin:2020ttj}. For young NSs with surface temperatures are in the range $T_{\text{sur}} \sim 10^5$--$10^6$~K, the bulk of the thermal emission occurs in the soft X-ray band, which has been observed by the Chandra X-ray Observatory~\cite{Jonker:2003au}. The ultraviolet counterpart of this emission is accessible to the Hubble Space Telescope~\cite{Kaplan:2011ay}. As the surface temperature drops to a few hundred Kelvin in older NSs, the thermal emission shifts predominantly to the infrared regime, which can potentially be within the observational capability of the future infrared telescopes~\cite{Chatterjee:2022dhp, Raj:2024kjq}. 
In this section, we explore relevant parameter space in the DM mass-coupling plane for relatively old, isolated neutron stars (INSs), considering different final observed surface temperatures. These temperatures provide valuable constraints on the DM content, the equation of state of dense standard matter, and the composition of both the stellar core and outer layers. In Sec.~\ref{sec:Evolution_NS_temperature}, we examined the thermal evolution of the NS interior, considering photon and neutrino emission as primary cooling mechanisms, and investigated the role of DM in heating, ultimately leading to a constant temperature. In our analysis, the NS mass and radius are critical parameters that determine the DM capture rate and the size of the thermalized DM core, thereby influencing the evolution of the DM population within the star. Additionally, in modeling the thermal evolution, the observed surface temperature depends on the surface gravitational acceleration $g_s$ and the gravitational redshift, both of which are functions of the NS mass and radius. For our analysis, we fix the NS mass and radius to be 1.4M$_{\odot}$ and 10 km respectively.   

\begin{figure}[t]
	\centering
	\includegraphics[scale=0.6]{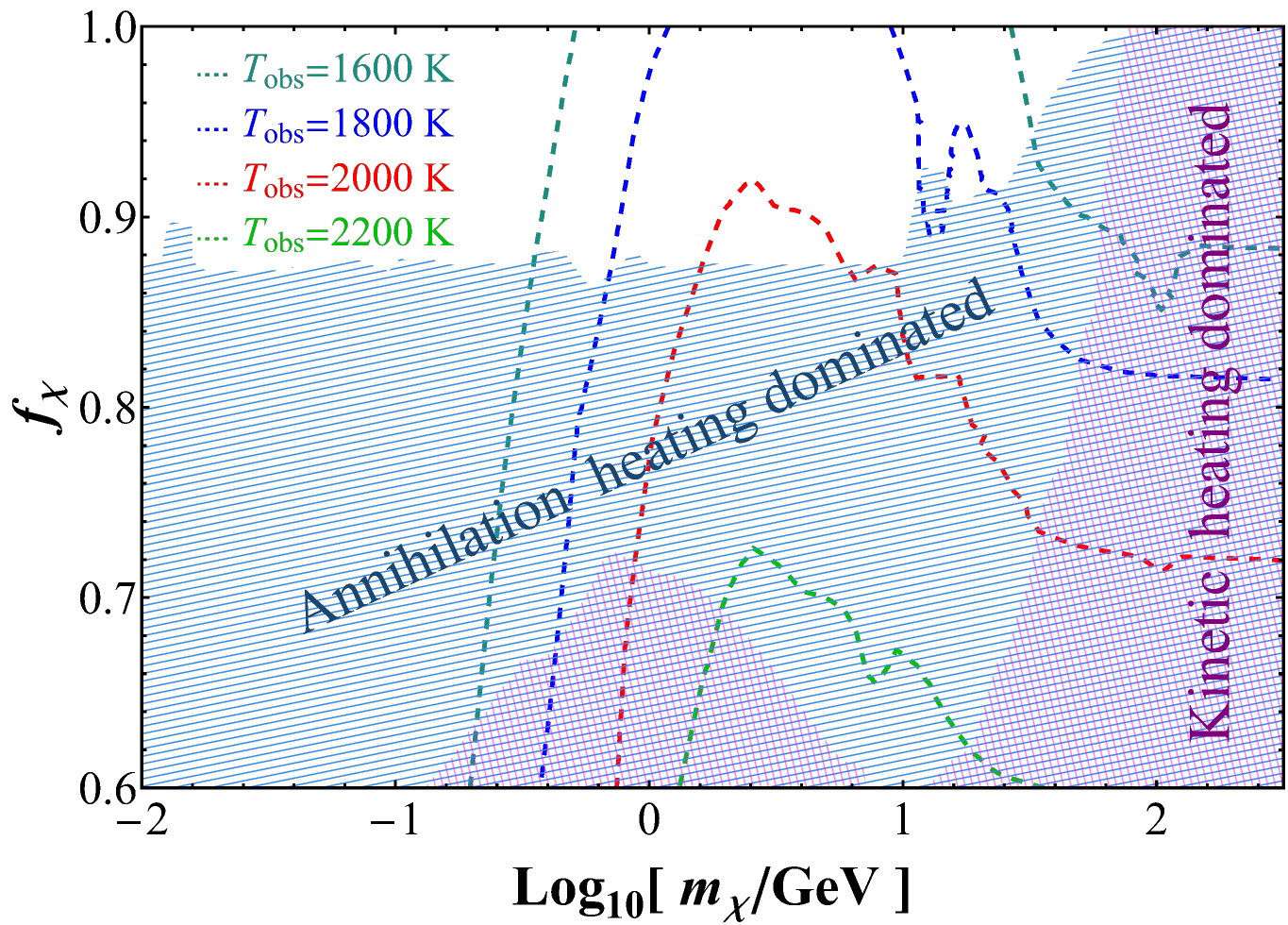}
	\caption{\small{Relevant parameter space for observational temperatures of  old NSs in the DM mass ($m_\chi$) vs initial DM particle fraction ($f_\chi$) plane, where the parameters are scaled to be dinemsionless. The contours of different temperatures (1600-2200 K) indicated by different coloured lines. In the blue-shaded region, heating due to DM annihilation dominates over cannibal heating and in the magenta-shaded region, kinetic heating is dominant. All the other relevant parameters of NS are taken to be same as Fig. \ref{fig:NS_Temperature_evolution}.}}
	\label{fig:NS_Temperature_Observation}
\end{figure}

We evaluate the thermal evolution of old INSs, accounting for heating due to DM self-annihilation via cannibal reactions and kinetic heating. For this section, we fix the $3\rightarrow2$ annihilation cross section by taking the trilinear and quartic self-coupling parameter to be $\mu=m_\chi/100~\text{and}~\lambda=1$ respectively. We solve the coupled evolution equations for the DM number density and the NS temperature, and evaluate the temperature at $t \sim 10^{9}$ years for various asymmetry parameters. The resulting contours indicate the combinations of $m_\chi$ and $f_\chi$ that yield specific observed temperatures which are shown in Fig. \ref{fig:NS_Temperature_Observation}. 
We find that in order for the surface temperature of old NSs to asymptotically settle around $\mathcal{O}(1000)$~K due to cannibal heating, the relevant DM mass lies around 100 MeV-100 GeV range. In the sub-GeV mass range, however, quantum corrections to the DM capture rate become significant. In the present work, we have not included the effects of Pauli blocking or Bose enhancement in the evolution of the captured DM population. Accounting for these quantum effects would render the capture rates temperature-dependent, and when self-consistently coupled to the temperature evolution, could lead to significant modifications of the viable parameter space. A detailed analysis incorporating these effects is left for future work~\cite{Dey2025}. The teal, blue, red and green contours correspond to observed NS temperatures of 1600, 1800, 2000 and 2200 K, respectively. As the temperature increases, the available parameter space capable of generating it becomes progressively smaller. 
In the blue-shaded region, heating due to $\chi\chi^\dagger \to \text{SM+SM}$ annihilation dominates over cannibal heating. On the other hand, in the magenta-shaded region, kinetic heating dominates over cannibal heating. This leaves a small viable region capable of producing observable temperatures potentially detectable by the JWST telescope \cite{Chatterjee:2022dhp, Baryakhtar:2017dbj}. Although JWST may be sensitive to blackbody temperatures of $\mathcal{O}(10^3)\,\mathrm{K}$ with exposure times of $10^5$--$10^6\,\mathrm{s}$, this sensitivity applies only to sources within distances of approximately $10\,\mathrm{pc}$ \cite{Raj:2024kjq}. However, pulsar population distributions inferred from local electron density maps show that such nearby targets are expected to be extremely rare \cite{Bramante:2024ikc}, making the observational prospects in JWST more challenging in practice.
Apart from JWST, there are several other upcoming telescopes that can potentially probe into this infrared regime of NS temperature. For example, InfraRed Imaging Spectrograph (IRIS) in Thirty Meter Telescope (TMT) can detect infrared signals between 0.8-2.4 $\mu$m, which can be translated for the possible spectrum of NS surface temperature between $\simeq1200-3600$ K \cite{Larkin_2016}. The observational sensitivity of Multi-AO Imaging Camera for Deep Observations (MICADO) instrument in Extremely Large Telescope (ELT) also encompasses similar range of spectrum \cite{Sturm_2024}. NS heating corresponding to a blackbody temperature of $T \simeq 2500\,\mathrm{K}$ could be detectable with the IRIS-TMT in the Y-band filter. For a source located at a distance of $50\,\mathrm{pc}$, an exposure time of $10^7\,\mathrm{s}$ would be required, while a source at $100\,\mathrm{pc}$ would necessitate exposures beyond $10^8\,\mathrm{s}$ \cite{Bramante:2024ikc}. Although such integration times are substantial for more distant targets, they are comparable to those achieved in ultra-deep field campaigns \cite{Illingworth:2013pda}. These considerations suggest that TMT could provide a viable future observational avenue for detecting NS heating in the temperature range illustrated in Fig.~\ref{fig:NS_Temperature_Observation}.

%%%%%%%%%%%%%%%%%%%%%%%%%%%%%%
\section{Summary and future prospects}
\label{sec:Summary}
%%%%%%%%%%%%%%%%%%%%%%%%%%%%%% 
In this work, we have explored the thermal evolution of neutron stars (NSs) in the presence of asymmetric dark matter (ADM) featuring number-changing self-interactions within the dark sector. Traditionally, ADM is considered to accumulate in significant numbers inside NSs due to the absence of efficient annihilation processes. We have proposed a modification to this picture by introducing cannibalistic self-interactions of the form $3 \rightarrow 2$, which allow for internal number depletion within the dark sector. This process results in a saturated DM population, preventing uncontrolled accumulation.
Additionally, these number-changing processes heat the DM sector, and the resulting energy is transferred to the stellar medium via DM-nucleon elastic scatterings. This additional heating alters the standard NS cooling trajectory, which is otherwise governed by neutrino and photon emissions, and can lead to significantly different surface temperatures in old neutron stars.
We have noticed that even a small initial asymmetry can be amplified over time, as annihilations rapidly deplete the antiparticle abundance, while the $3 \rightarrow 2$ process can regenerate a non-zero antiparticle density even in a fully asymmetric scenario. The resulting thermal impact depends sensitively on the degree of asymmetry: larger asymmetry enhances the total accumulated DM through an increased effective capture rate, thereby strengthening cannibal heating, while simultaneously suppressing annihilation heating due to the reduced antiparticle abundance. We find that cannibal heating can dominate over annihilation heating in the high-asymmetry regime, establishing $3\rightarrow2$ process as a previously overlooked mechanism of NS heating.

To quantitatively investigate this effect, we solved the coupled evolution equations for the DM number and stellar temperature, taking into account realistic NS parameters. We demonstrated that the observable surface temperatures of the old isolated neutron stars can be produced within this framework for appropriate choices of the DM mass $m_\chi$ and initial DM particle relative abundance $f_\chi$. Furthermore, the dominance of annihilation and kinetic heating over the cannibal process indicates viable values for both parameters.
Our results suggest that thermal observations of old NSs can serve as valuable probes of ADM properties, offering complementary constraints on DM mass and couplings. An important extension of this work would be to incorporate proper quantum statistical effects in the dark matter capture rate. In particular, accounting for Fermi-Dirac or Bose-Einstein statistics can introduce a temperature dependence in the capture process, which may influence the early-time behaviour of DM accumulation in neutron stars. 
We note that this work highlights the potential of stellar astrophysics to illuminate the nature of dark matter through its subtle yet observable impact on compact objects.

\acknowledgments{We thank Tarak Nath Maity for helpful comments on the manuscript. UKD acknowledges support from the Anusandhan National Research Foundation (ANRF), Government of India under Grant Reference No.~CRG/2023/003769. SG acknowledges the support of postdoctoral fellowship from IISER Berhampur.}

%%%%%%%%%%      References     %%%%%%%%%%%%%
\bibliographystyle{JHEP}
\bibliography{ref.bib}
%%%%%%%%%%%%%%%%%%%%%%%%%%%%%%%%

\end{document}